# All-optical wavelength-tunable narrow-linewidth fiber laser

**Yujia Li, Tao Zhu*, Shihong Huang, Lei Gao, Tianyi Lan, and Yulong Cao**

*Key Laboratory of Optoelectronic Technology & Systems (Ministry of Education), Chongqing University, Chongqing 400044, China*

**Corresponding author: zhutao@cqu.edu.cn*

## Abstract

Parameter regulations of narrow-linewidth fiber lasers in frequency domain has drawn considerable interests for widespread applications in the light quantum computing, precise coherent detection, and generation of micro-waves. All-optical methods provide compact, precise and fast accesses to achieving these lasers with wavelength-tunability. Here, the optical-thermal effects of graphene is utilized to precisely control operations of free-running lasers with a tuning speed of 140 MHz/ms. Assisted by the single-longitude-mode operation and linewidth suppression of stimulated Brillouin backscattering, we obtain an optical-controllable ~750 Hz fiber laser with a wavelength-tuning range of 3.7 nm.

## 1. Introduction

Narrow-linewidth fiber lasers with flexible output performances have attracted widespread interests because of their advantages, such as long coherent length, excellent spectral purity and compatibility with other fiber devices for easiness [1-4]. Considerable researches have been directed toward pursuing transplantable, precise and convenient wavelength-tuning methods for those lasers. Recently, the main tuning methods are based on three kinds of external physical filed: additional mechanical stress, electric- and acoustic-field derive [5-9]. Due to the limit of the stepping distance (larger than ~1 um) of common mechanical adjusting systems, it is quite difficult to ensure the consistency of tuning tendency and achieve a high tuning precision, especially for manipulated tunable fiber devices with geometrical sizes of the micrometer order. The tuning methods based on electric and acoustic-field require complex and expensive drive setups, and the energy couple and transfer between different physical field is unavoidable, which goes against the all-optical integration with other optic devices.

Contrasted with those methods, all-optically tuning lasing wavelengths owns unique advantages, such as compact implementation, easy operation and low cost. The optic nonlinearities, such as Raman self-frequency shift and selective parametric process, have been tried achieving all-optical wavelength-tuning [10-11]. Considering fiber' s weak nonlinearity, both of significant pump power and properly designed fiber length are essential. Therefore, those methods are just suitable for the ultrafast laser systems, where the pulse laser can provide ultrahigh peak power. Furthermore, energy transfer exited in the nonlinear process induces an ineluctable crosstalk to the controlled laser. Thus, to achieve inherent and transplantable optical-controlling of lasing wavelengths, the controlling light has to be an independent part of laser systems, and only regulates physics of devices inside laser systems.

Graphene, regarded as one of the most remarkable novel 2D materials, can perform diverse properties, such as the saturable absorption [12-16], Kerr effect [17-18], photo-thermal effect [19, 20], etc. The flexibility of single-layer graphene makes it easily integrated with various fiber devices to modulate their physics by an additional controlling light. Therefore, graphene with a carrier based on micro-nano fiber devices can become an excellent medium of interaction between the controlling light and manipulated laser. Up to now, some ultra-fast all-optical modulation fiber devices based on graphene with a response speed of nanosecond order has been achieved by stimulating the ultrafast nonlinearity of graphene. Nevertheless, geometrical sizes of those fiber devices has to be controlled near ~1 um [17]. Different from just acting as an optical modulation device, the larger insertion loss has to be taken into consideration for their applications in narrow fiber laser systems. To make fiber devices more robust and lower insertion loss, FBGs can be optical-tuned through the strong optical-thermal effect of graphene, where the transverse sizes of devices can reach tens of micrometers, resulting in a more convenient fabrication [20]. Moreover, the pure fiber structure and steerable bandwidth make this device compatible with all-fiber tunable narrow fiber lasers for easiness. However, the previous works mainly emphasis on studies of statics and dynamics of those all-optical devices.

In this work, an optical-controlled fiber filter, fabricated by coating graphene on the surface of the mico-FBG, is employed in the wavelength-tuning of narrow linewidth fiber laser. Due to the weak evanescent wave interaction between FBG and graphene, reflected wavelengths can be tuned by an external controlling pump light. The optical-controlled devices is inserted into a double cavity fiber laser cavity to modulate the lasing mode competition and achieve tunable output wavelength. Due to the low propagation loss and uniform index distribution of fibers with a long distance, the stimulated Brillouin scattering (SBS) can be stimulated and accumulated for easiness with enough pump light, which can be used as a linewidth compression component for the narrow bandwidth of tens of Megahertz [21-23]. Then this optical-controlling laser, used as the pump, is injected into the polarization-maintaining cavity to stimulate the SBS laser and achieve the mode-selection and linewidth compression. Here, by the combination of the all-optical fiber device and SBS based laser linewidth compression, the optical-regulation of lasing wavelengths are achieved, which provides an innovate avenues for the parameter controlling for narrow-linewidth fiber laser systems.

## 2. Devices fabrication, principle and test

The uniform FBG used in our experiment, fabricated by the ultraviolet engraving, has the same transverse geometrics with the standard single-mode fiber (Corning, 28e). For the purpose of ensuring a effective contact between graphene and evanescent wave, the MFBG with a diameter of 15.6 um is prepared by etching a FBG in the hydrofluoric acid with a concentration of 8% for 6 hours and 40 minutes to reduce part of cladding. Thus, the periodical index distribution in core is not influenced. As shown in Fig. 1(a), the length of the gating region is 12 mm, including a graphene-coated region. A loss of ~2 dB is caused during the etching process for the light escaped from the cladding, shown in Fig. 1(b). After the MFBG etched, the single-layer graphene can be conveniently transferred to the surface of MFBG' s cladding in deionized water. A longer graphene region is propitious to the optical-tuning sensitivity. However, considering that this device can be compatible with fiber laser systems, the total loss is controlled to ~6 dB by properly optimizing the length of graphene-coated region as 5.8 mm. The graphene-induced change of index causes a wavelength-shift of merely 0.1 pm. Therefore, wavelength-shifts during fabrication process is mainly due to the combined influences of changed mechanical stress.

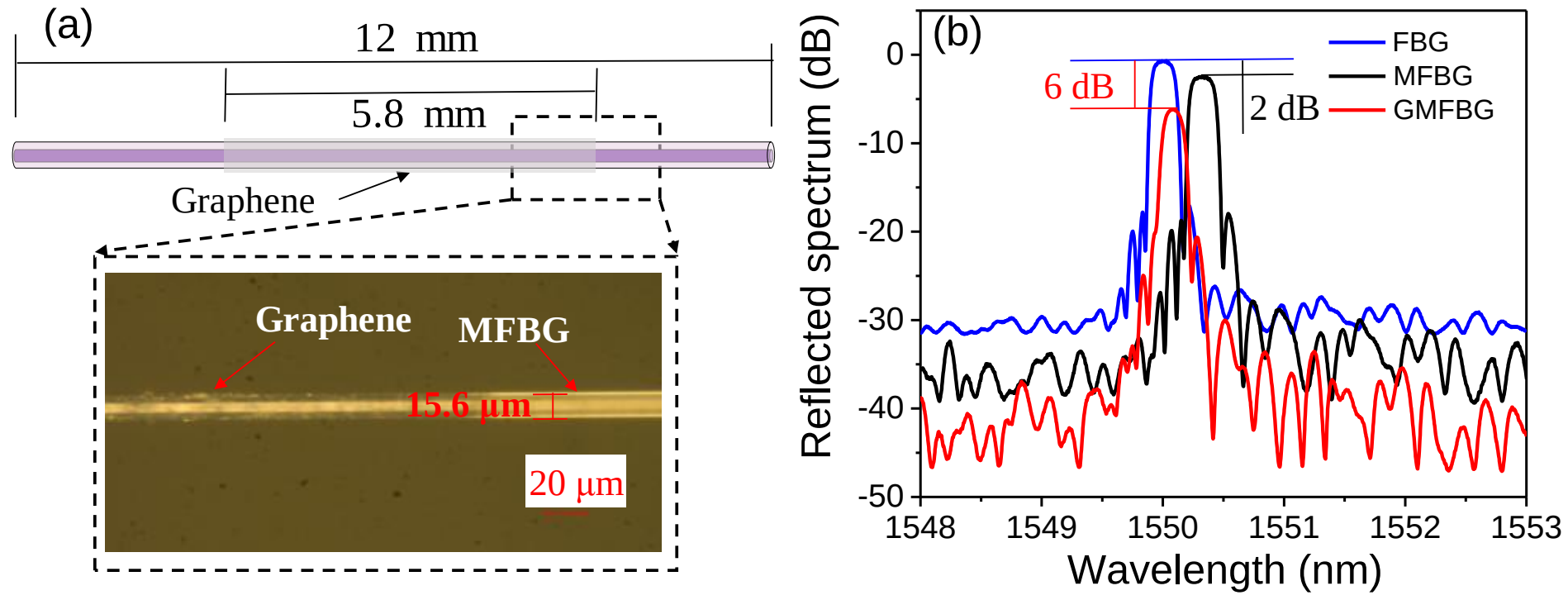

Fig. 1 (a) Geometrics of the fabricated GMFBG; (b) reflected spectra of the FBG, MFBG and GMFBG.

Figure 2(a) shows the interaction between graphene and MFBG. When pump light is entering the MFBG, only evanescent wave escaped from cladding has a contact with graphene. Graphene with a single-layer structure is thermally damaged for easiness when used in laser systems. To ensure this device can withstand a pump power of several-hundred milliwatt and obtain a wider tuning range, we theoretically estimate the relative intensity of the evanescent wave through the energy distribution of the fundamental mode at 980 nm on the cross section of the graphene-coated MFBG (GMFBG), as shown as Fig. 2(b). The inset with a red box shows the enlargement of the localized region. The ratio of the intensity at the boundary of the cladding to that at the center core is ~$10^{-6}$. For instance, when a light with a power of even 1 W is injected into the GMFBG, the light contacting with graphene has a power of only ~1 uW, which is hardly possible to cause the thermal damage of graphene.

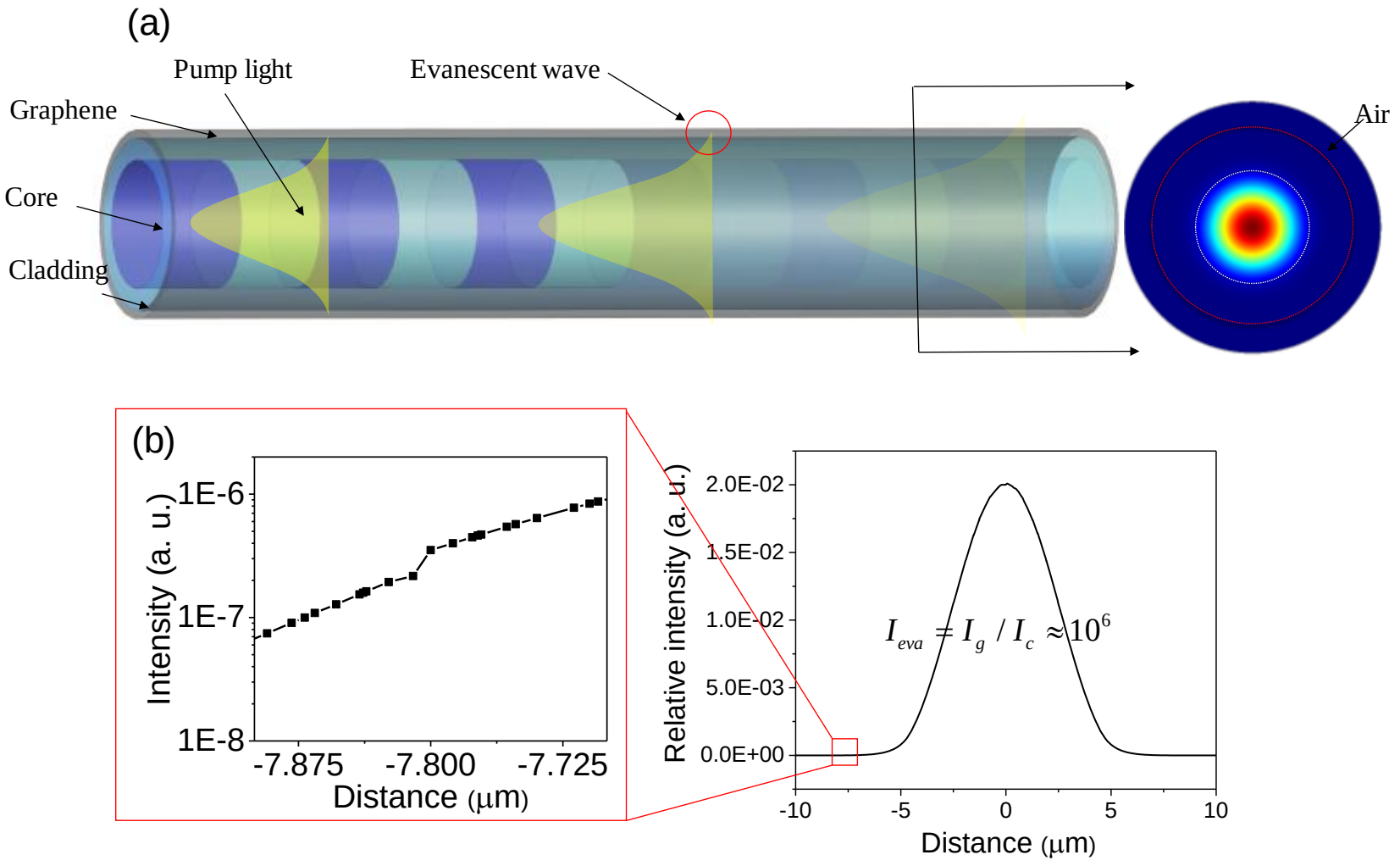

Fig. 2. (a) Evanescent wave interaction between MFBG and graphene; (b) mode distribution of GMFBG’ s cross section.

The static optical-tuning performance of the GMFBG is tested by the system shown as Fig. 3(a). The BBS (broadband source) and the circulator is used to observe the reflected spectrum. The controlling light with a wavelength at 980 nm is far from the reflected center wavelength, and it will pass the GMFBG and be filtered by the circulator that operates at C-band. By enhancing the power of the controlling light coupled into GMFBG, the reflected spectrum has a red-shift, as seen as Fig. 3(b). The evanescent wave of 980 nm light exposed on the graphene will stimulate the graphene’ s optical-thermal effect and generate significant Joule heating, resulting in the change of graphene’ s index. The reflected center wavelength of GMFBG is depended on the effective index. There are mainly two factors to cause the change of the mix-waveguide’ s effective index. The higher temperature of graphene will enlarge the index of itself, resulting the whole device’ s larger index. On the other hand, the considerable Joule thermal of graphene also heats the MFBG. The red-shift of the wavelength appears, companied with the adverse deterioration of reflected spectrum. This is mainly due to the non-uniform distribution of graphene, as the gating region is not entirely covered by graphene. Thus, the periodical index distribution of the original MFBG is destroyed. The longer graphene-coated region is in favour of the improvement of the tuning sensitivity and spectral quality. However, the larger insertion loss has to be taken into consideration for the strong absorption of graphene. The height of the side-lope is not beyond the 1/3 of the peak wavelength. Thus, the side-lope will be submerged though the mode competition when used in tuning laser systems.

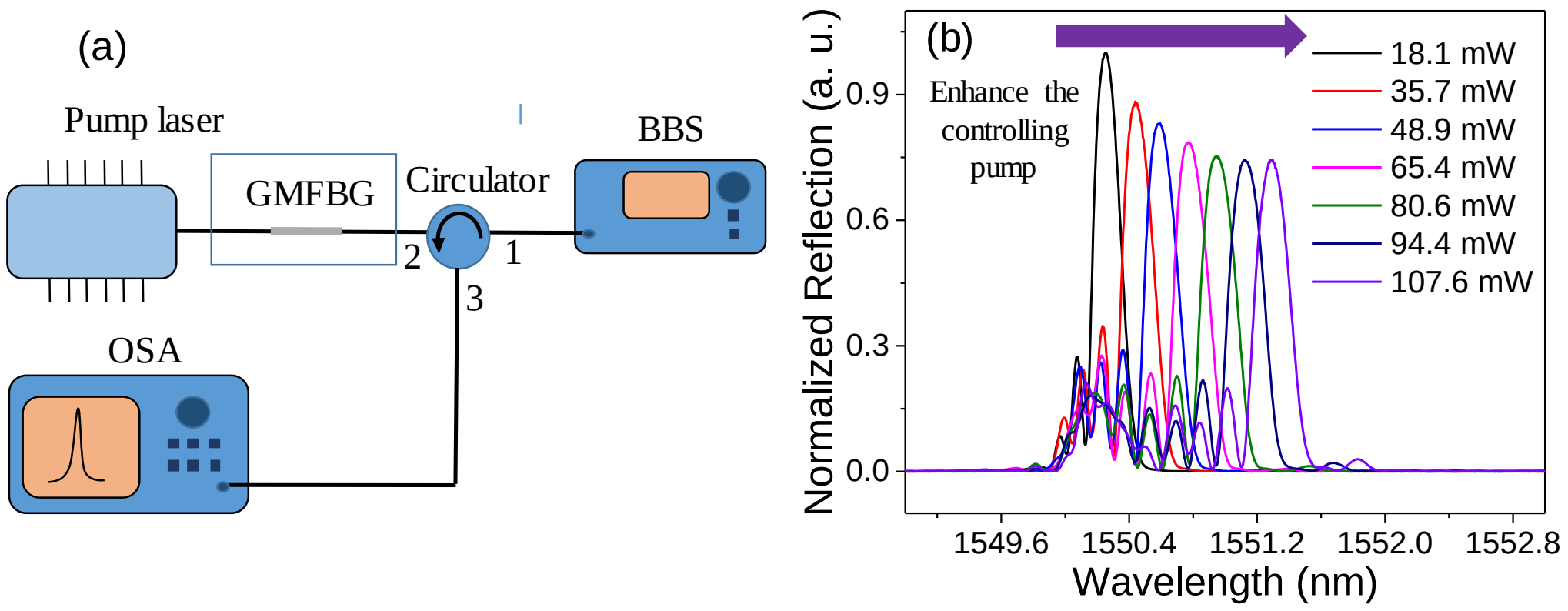

Fig. 3 Test of GMFBG' s static characteristics. (a) Experimental setup of the test system; (b) normalized reflected spectra with the enhancement of the controlling pump light.

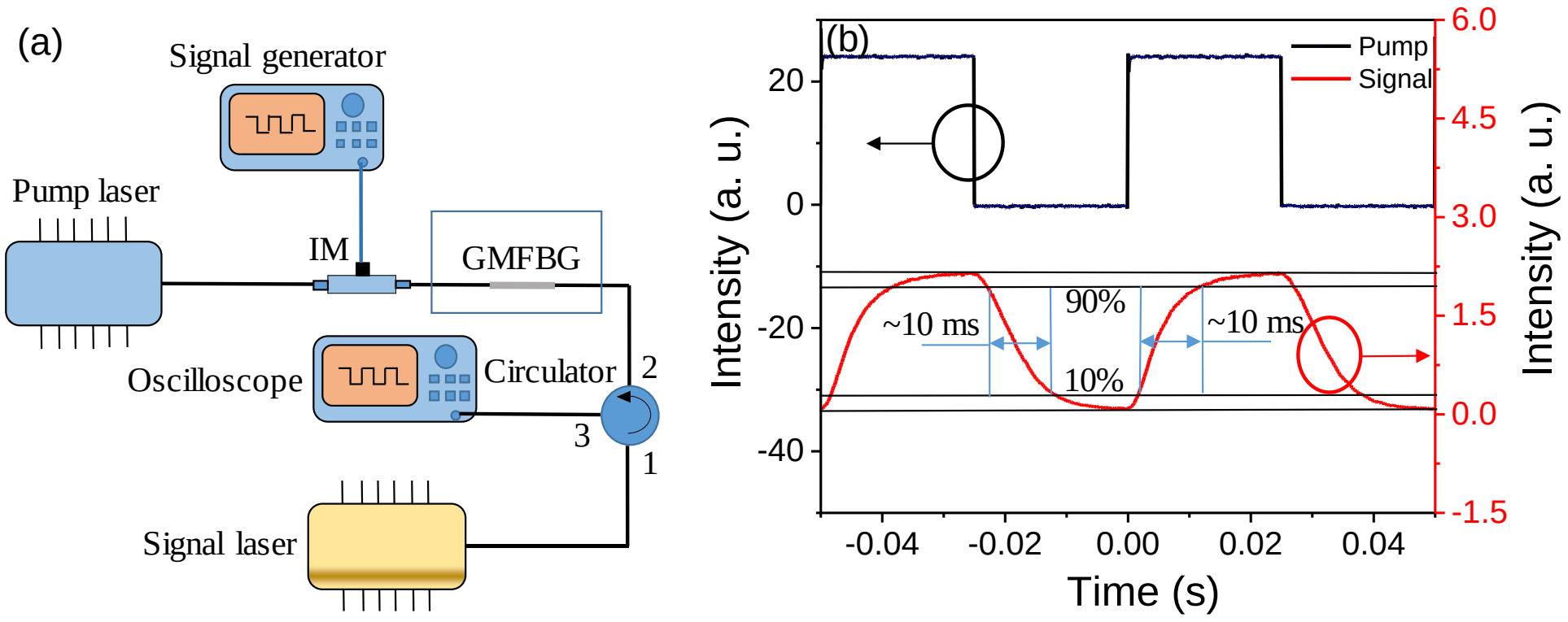

Fig. 4 Test of GMFBG' s dynamic characteristics; (a) experimental setup of the test system; (b) comparison of pump light and signal light.

Fig. 4(a) shows the system for the test of the dynamics of the GMFBG. We use a signal laser with a wavelength aligning the falling edge of the reflected spectrum, when no controlling light is injected. The controlling light is modulated as a square wave in the time domain. The controlling light intensity converts between the high level (1 mW) and low level (0 mW) with the modulation frequency of 20 Hz, which causes an extinction ratio of ~20 dB to the signal light, shown as Fig. 4(b). The response time is estimated with the signal laser changing from 10% to 90%, and the rising and falling times are ~10 ms, which is corresponding to a tuning speed of ~140 MHz/ms, considering a wavelength shift of 0.014 nm of the GMFBG induced within a modulation period. The dynamics of this device also determines the tuning speed of the following narrow-linewidth laser. This can be further improved by a smaller diameter of MFBGs for the ultrafast nonlinear absorption of graphene. If the diameter of FBG after etched is too close to core, or smaller than the core's diameter, the spectral quality will be seriously reduced, together with a larger insertion loss. This graphene-assisted all-optical tuning method can be also transplanted to other higher Q-factor micro-nano devices, especially for whispering gallery micro-cavities that have been widely applied in the linewidth compression for ultra-narrow linewidth lasers.

## 3. Experimental setup

The fast optical-controlled performance of the GMFBG is employed in the wavelength controlling for the narrow-linewidth fiber laser shown as Fig. 5(a). The whole system are consisted of three parts: optical-controlling wavelength-tunable laser, Brillouin laser and measurement system. The first part is used to generate a wavelength-tunable laser as the pump of the following Brillouin laser. The 980 nm pump laser is coupled into the cavity by a 980/1550 WDM (wavelength division multiplexer). We use 1 m EDF as the gain medium, and a polarization controller inside the cavity is necessary for the polarization-sensitivity of Brillouin laser. The Isolator is used to determine the operation direction of the laser. The GMFBG is connected with the cavity by the circulator 1. Another 980 nm pump laser is essential to optically tune the GMFBG. This controlling pump light cannot enter into this cavity, and will be filtered by the circulator. Therefore, it just controls GMFBG's physics without any crosstalk to the laser inside the cavity. Here, we use OC1 (coupler 1) and OC2 to build a double-cavity structure to enlarge the FSR (free spectrum range) and enhance the stability. The feedback and output of the laser cavity is achieved by OC3. The length of these two cavities are ~13 m and ~14 m. The coherent FSR of laser at output 1 is ~200 MHz, which is the least common multiple of the two cavities' FSRs.

The SBS has a significant compression effect for the laser linewidth, and it is very easy to achieve the SLM operation with a fiber cavity of less than 20 m for the narrow Brillouin gain bandwidth (~20 MHz) [23]. In conventional fibers, we use the backscattering light mainly including the stokes-light. When the pump light is injected into the fiber cavity, the backscattering stokes light is reversely propagated in the cavity. Once the pump light intensity satisfying the condition of the laser oscillation, the excited stokes photon can reduce the phase noise of laser for the slow attenuation, resulting a narrower linewidth than the pump light. The Er-doped fiber amplifier （EDFA） located behind output 1 is used to amplify the optical-tunable laser and stimulate enough Brillouin gain condition. To ensure a SLM operation and enough Brillouin gain to form the laser, the Brillouin laser cavity is consisted of 20 m polarization-maintaining fiber and two polarization-maintaining fiber devices (circulator 2 and OC4), as shown as the part Ⅱ. The used total cavity length is calculated by the equation:

$$L \cong 2c / n\Delta f_B,$$

where c is the light speed in vacuum, n is the index of fiber, and $\Delta f_B$ ,defined as the bandwidth of Brillouin bandwidth, is ~20 MHz. This value of $\Delta f_B$ has approximated the intrinsic limitation of Brillouin gain in fiber. Here, the all-polarization-maintaining cavity is used to enhance the stability and reduce the frequency noise of the output laser for the strong polarization-sensitivity of Brillouin gain.

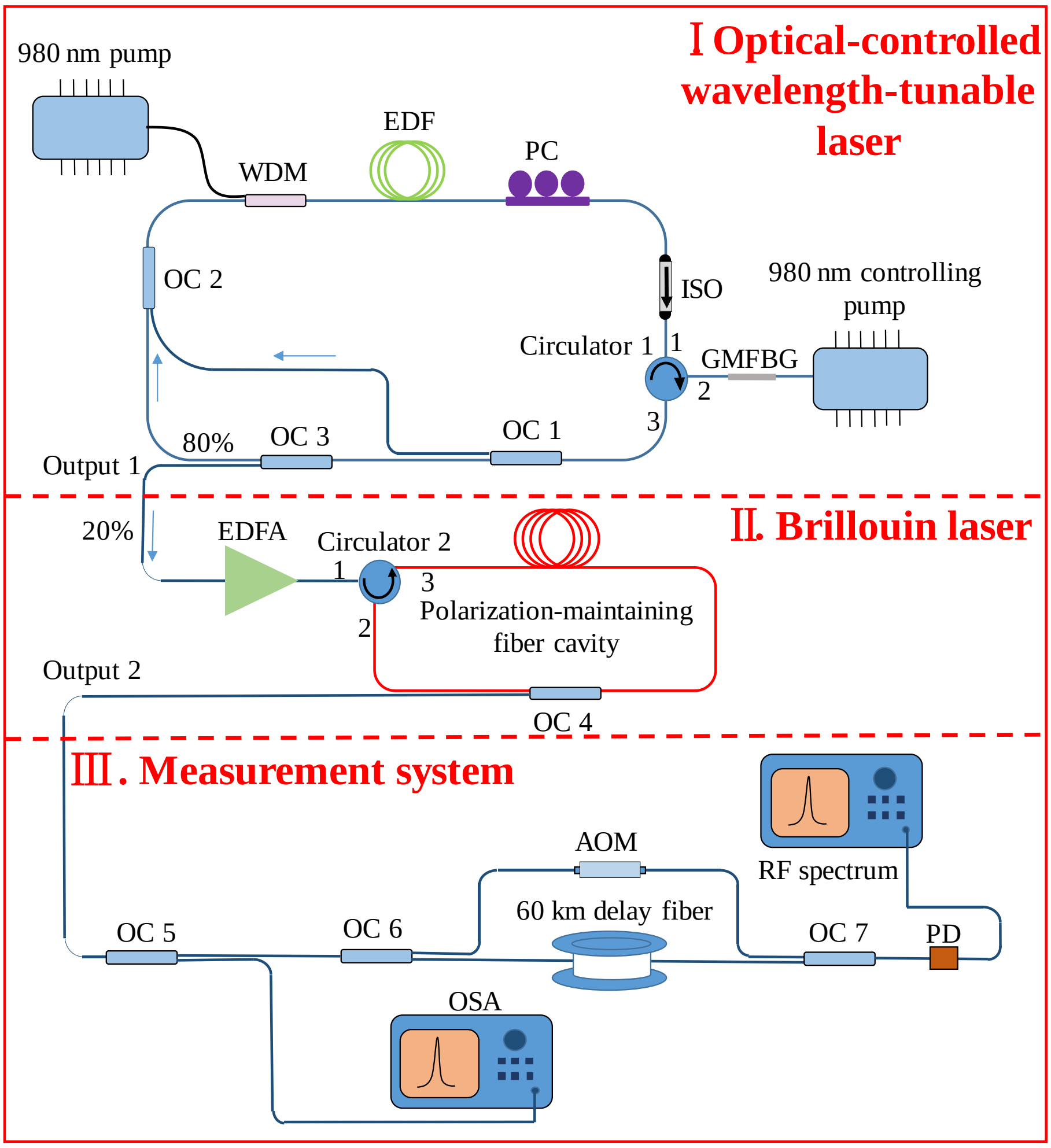

Fig. 5 Experimental setup of the optical-controllable wavelength-tunable fiber laser and the measurement system of spectra and linewidths.

The third part is used to simultaneously measure spectra by an optical spectrum analyzer (OSA, AQ6370D, YOKOGAWA) with a resolution of 0.02 nm and linewidths by the delay self-heterodyne method of the output laser. The acoustic-optic modulator in the up-arm of the Mach-Zehnder interferometer is used to generate a frequency-shift of 80 MHz, and the 60 km delay standard single-mode fiber can induce a delay of 300 μs for to the light in the other arm. The beat frequency signal is generated within the PD (DET01CFC, Thorlabs) with a bandwidth of 2 GHz, which is observed by a radio-frequency (RF) analyzer with a bandwidth of 30 GHz.

## 4. Experimental results and discussion

In this work, output wavelengths of free-running pump laser (part Ⅰ) is controlled by the GMFBG with an additional 980 nm pump light. Fig.

6(a) shows the wavelength tuning process of the pump laser (output 1). With the increment of the controlling pump power, the pump laser has a red-shift, owing to the all-optical performance of the GMFBG. As the previous discussion, the spectral deterioration of the GMFBG has no influence on the output laser for the coherent mode competition during the oscillation process. For further mode-selection and linewidth compression, this tunable pump laser after amplified by an ( EDFA ) is injected in the polarization-maintaining Brillouin cavity. To obtain a stable Brillouin laser, we properly adjusting the polarization controller inside the tunable pump laser cavity until the power at output 2 reaches a maximum value. The power of the amplified laser is controlled by adjusting the driving current of the EDFA, and the oscillation threshold of the Brillouin laser is about 90 mW.

When we set the output power of the EDFA as 169 mW, the power of the Brillouin laser (output 2) is 27 mW and has a ~0.084 nm wavelength-shift about the pump laser, which is also well coincided with the theoretical frequency-shift of Brillouin laser, as seen as the inset within a black dashed box in Fig. 5(b). For the low phase-noise characteristics of oscillated Brillouin laser, the contrast ratio of the Brillouin laser is improved by over 20 dB, compared with the pump laser. Because the frequency-shift between the tunable pump laser and the Brillouin laser is a constant value (~11 GHz) with the Brillouin cavity well fixed at a stable temperature in lab conditions, the laser at output 2 should have the same tuning performance as that at output 1. For the proper cavity length optimization of the Brillouin cavity and narrow Brillouin gain bandwidth, the output laser can operate at a SLM state for easiness. Figure 6(c) shows the RF spectrum at output 2, measured by the delay self-heterodyne system (part Ⅲ). The main peak is positioned at 80 MHz, which equals the AOM’ s modulation frequency. There are no side-modes appearing within a frequency span of 1 GHz, indicating a good SLM operation of the laser. Figure 6(d) shows the corresponding RF spectrum evolution with a the variation of EDFA’ s output power, demonstrating that this cavity always keeps a good SLM operation under different pump powers, considering the coherent FSR of the Brillouin cavity is ~10 MHz.

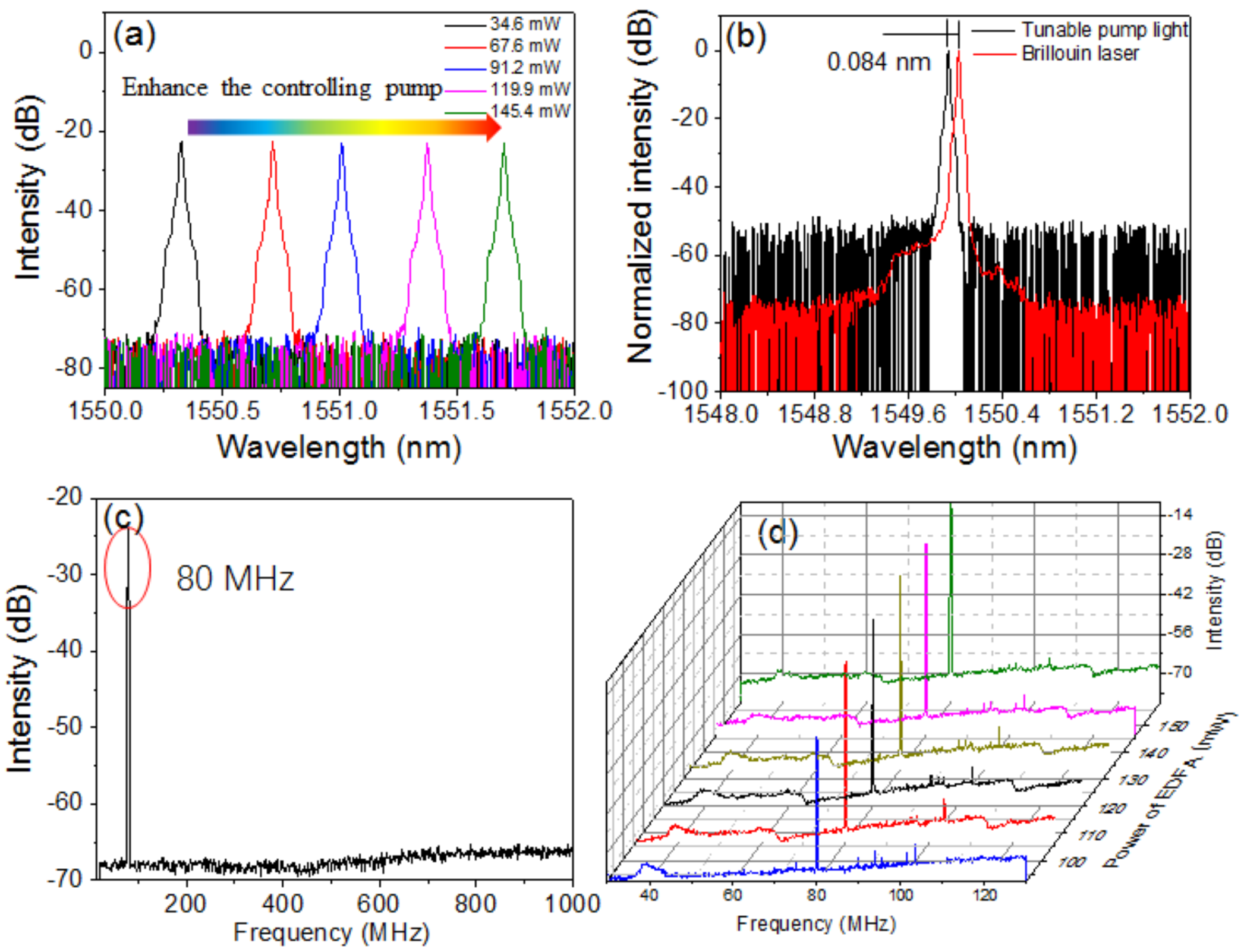

Fig. 6 (a) Output spectra at output 1 under different controlling pump powers; (b) spectra at output 1 and the Brillouin laser. (c) RF spectrum of the Brillouin laser with a frequency span of 1 GHz; (d) measured frequency noise with a EDFA' s power of 160 mW; (d) corresponding color-coded RF spectra at output 2 with a frequency span of 100 MHz measured by the delay self-heterodyne method.

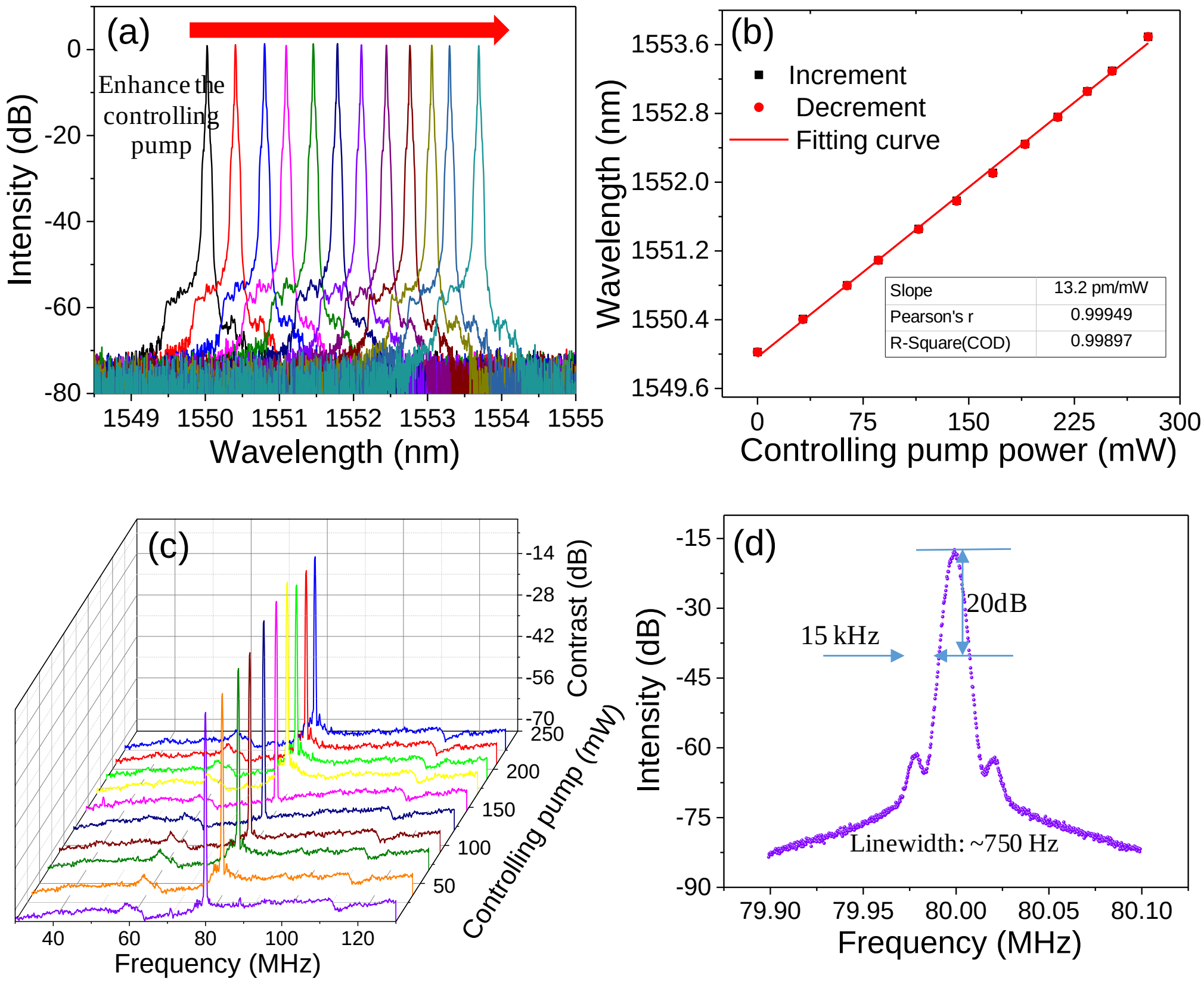

Fig. 7 (a) Output spectra at output 2 with the enhancement of the controlling pump laser; (b) wavelength positions with the corresponding controlling pump power; (c) RF spectra under different controlling pump powers with a frequency span of 100 MHz; (d) RF spectra with a frequency span of 200 kHz with the controlling pump power of 67.6 mW.

We keep the output power of the EFDA as 169 mW unchanged and only adjust the controlling pump intensity. Fig. 7(a) shows that the spectrum of the Brillouin laser has a red-shift with the enhancement of the controlling pump laser, indicating a coincident tuning tendency with the tunable pump laser. The controlling pump just contacts with the GMFBG and is filtered by the circulator. Therefore, there is no direct energy coupling between the Brillouin laser and controlling pump light. The corresponding peak wavelength is linearly changed within a tuning range of 3.7 nm with the controlling pump varies from 0 mW to 275.4 mW with a tuning sensitivity of ~13.2 pm/mW during the tuning processes with the increment and decrement of powers, as seen in Fig.8 (b). The linear $R^2$ values of two processes are over 0.999897, indicating a stable and precise tuning performance. The GMFBG is tuned within the maximum power range we can provide and not be thermally damaged for a weak evanescent wave interaction as the previous discussions. Moreover, the change of the wavelength always keeps a good linear tendency during the while tuning process. Therefore, the tuning range can be further widened by the higher controlling pump power. The tuning precision is dependent on the minimum adjusting stepping and stability of the controlling pump laser, owing to the sensitive optic-thermal performance of graphene and good package for the GMFBG. To approximate a higher tuning precision, the Q-factor (~$10^4$ for FBGs) of the optical-filter has to be improved , for example, utilizing the ultra-finesse Whispering-gallery-mode (WGM) resonators. The tuning speed is mainly limited by the GMFBG's dynamic response time of ~10 ms that is much longer than the oscillation relaxation time of population inversion in Er-dope fiber. Figure 8(c) shows the RF spectra with a frequency span of 100 MHz under different controlling pump powers, indicating that the output laser always keeps a SLM operation at all wavelength channels. The Brillouin laser's 20-dB linewidth of 15 kHz is achieved during the tuning proves, as seen as Fig. 7(d), corresponding to a estimated practical linewidth of 750 Hz.

## Conclusion:

We have induced an all-optical wavelength-tuning method to narrow-linewidth fiber laser systems. Owing to the weak evanescent wave interaction between graphene and the mico-FBG, the optical-controlled can withstand a pump power of hundreds of milliwatt. By controlling an additional pump power injected into the GMFBG, the wavelengths of the narrow linewidth Brillouin fiber laser can be linearly tuned with a sensitivity of 13.2 pm/mW. Linewidths of ~750 Hz are achieved during the whole tuning process, due to the linewidth compressing effect of Brillouin laser. Considering unique advantages of all-optical controlling, including stable, low cost and fast, such a fiber laser has tremendous potential applications, such as generation of micro-wave, precision sensing and communications. Furthermore, due to the transportation of this inherent tuning method, this optical-controlled idea can be also popularized in other ultra-fast laser systems.

## Funding:

This work was supported by the Key Research and Development Project of Ministry of Science and Technology (2016YFC0801200), the National Natural Science Foundation of China (NFSC) (61635004, 61405020, 61520106012, 61705023), the National Postdoctoral Program for Innovative Talents (BX201600200), the Science Fund for Distinguished Young Scholars of Chongqing (CSTC2014JCYJJQ40002).

## Reference:

1. J. G. Tang and J. Q. Sun, "Stable and widely tunable wavelength spacing single longitudinal mode dual-wavelength erbium-doped fiber laser," Opt. Fiber Technol. 16(5): 299–303 (2010).
2. M. Choma, M. Sarunic, C.Yang, and J. Izatt, "Sensitivity advantage of swept source and Fourier domain optical coherence tomography," Opt. Express 11(18): 2183–2189 (2003).
3. T. Zhu, F. Chen, S. Huang and X. Bao, "An ultra-narrow linewidth fiber laser based on Rayleigh backscattering in a tapered optical fiber," Laser Phys. Lett. 10, 055110 (2013).
4. T. Zhu, B. Zhang, L. Shi, M. Deng, J. Guo, and X. Li, "Tunable dual-wavelength fiber laser with ultra-narrow linewidth based on Rayleigh backscattering", Opt. Express 24(2): 1324–1330 (2016).
5. M. W. Maeda, S. Patel, D. A. Smith, C. Lin, M. A. Saifi, and A. V. Lehman, "An electronically tunable fiber laser with a liquid-crystal etalon filter as the wavelength-tuning element," IEEE Photon. Technol. Lett. 2(11):787-789 (1990).
6. Y. W. Song, S. A. Havstad, D. Starodubov, Y. Xie, A. E. Willner, and J. Feinberg, "40-nm-wide tunable fiber ring laser with single-mode operation using a highly stretchable FBG", IEEE Photon. Technol. Lett. 13(11):1167-1169 (2001).
7. G. A. Ball and W W Morey, "Continuously tunable single-mode erbium fiber laser," Opt. Lett. 17(6): 420–422 (1992).
8. D. A. Smith, M. W. Maeda, J. J. Johnson, J. S. Patel, M. A. Saifi, and A. Von Lehman, "Acoustically tuned erbium-doped fiber ring laser," Opt. Lett. 16(6): 387–389 (1991).
9. L. Huang, X. Song, P. Chang, W. Peng, W. Zhang, F. Gao, F. Bo, G. Zhang, and J. Xu, "All-fiber tunable laser based on an acousto-optic tunable filter and a tapered fiber," Opt. Express 24(7): 7449-7455 (2016).
10. N. Nishizawa and T. Goto, "Compact system of wavelength-tunable femtosecond soliton pulse generation using optical fibers," IEEE Photon. Tech. Lett. 11(3): 325–327 (1999).
11. P. Wei, C. Li, C. Zhang, J. Wang, Y. Leng, and R. Li, " High-Conversion-Efficiency and Broadband Tunable Femtosecond Noncollinear Optical Parametric Amplifier," Chin. Phys. Lett. 25(7): 2514-1517 (2008).
12. G. K. Lim, Z. L. Chen, J. Clark, R. G. S. Goh, W. H. Ng, H. W. Tan, R. H. Friend, P. K. H. Ho, and L. L. Chua, "Giant broadband nonlinear optical absorption response in dispersed graphene single sheets," Nat. Photon., 5: 554–560 (2011).
13. Z. Luo, M. Zhou, J. Weng, G. Huang, H. Xu, C. Ye, and Z. Cai, "Graphene-based passively Q-switched dual-wavelength erbium-doped fiber laser," Opt. Lett. 35(21): 3709–3711 (2010).
14. H. Zhang, D. Y. Tang, L. M. Zhao, Q. L. Bao, and K. P. Loh, "Large energy mode locking of an

erbium-doped fiber laser with atomic layer graphene," Opt. Express 17(20): 17630–17635 (2009).
15. C. Gao, L. Gao, T. Zhu, and G. Yin, "Incoherent optical modulation of graphene based on an in-line fiber Mach–Zehnder interferometer," Opt. Lett. 42(9):1708–1711 (2017).
16. Q. Wen, W. Tian, Q. Mao, Z. Chen, W. Liu, Q. Yang, M. Sanderson, and H. Zhang, "Graphene based All-Optical Spatial Terahertz Modulator," Sci. Rep. 4: 7409 (2014).
17. S. Yu, X. Wu, K. Chen, B. Chen, X. Guo, D. Dai, L. Tong, W. Liu, and Y. Ron Shen, "All-optical graphene modulator based on optical Kerr phase shift," Optica, 3(5):541–544 (2016).
18. R. Nandkishore and L. Levitov, "Polar Kerr effect and time reversal symmetry breaking in bilayer graphene," Physic. Rev. Lett. 107, 097402 (2011).
19. X. Gan, C. Zhao, Y. Wang, D. Mao, L. Fang, L. Han, and J. Zhao, "Graphene-assisted all-fiber phase shifter and switching," Optica 2(5): 468–471 (2015).
20. X. Gan, Y. Wang, F. Zhang, C. Zhao, B. Jiang, L. Fang, D. Li, H. Wu, Z. Ren, and J. Zhao, "Graphene-controlled fiber Bragg grating and enabled optical bistability," Opt. Lett. 41(3): 603–606 (2016).
21. S. Huang, T. Zhu, G. Yin, T. Lan, L. Huang, F. Li, Y. Bai, D. Qu, X. Huang, F. Qiu, "Tens-of-Hz narrow-linewidth laser based on stimulated Brillouin and Rayleigh scattering, " Opt. Lett. 2017.
22. A. Debut, S. Randoux, and J. Zemmouri, "Linewidth narrowing in Brillouin lasers: Theoretical analysis," Phys. Rev. A. 62(2), 023803 (2000).
23. A. Debut, S. Randoux, and J. Zemmouri, "Experimental and theoretical study of linewidth narrowing in Brillouin fiber ring lasers," J. Opt. Soc. Am. B 18(4): 556–567 (2001).